\documentclass[preprint]{elsarticle} 
\usepackage[hyphens]{url}

  \journal{Ecology Letters} 

\usepackage{lineno} 
\providecommand{\tightlist}{%
  \setlength{\itemsep}{0pt}\setlength{\parskip}{0pt}}

\usepackage{graphicx}

\usepackage[T1]{fontenc}
\usepackage{lmodern}
\usepackage{amssymb,amsmath}
\usepackage{ifxetex,ifluatex}
\usepackage{fixltx2e} 
\IfFileExists{upquote.sty}{\usepackage{upquote}}{}
\ifnum 0\ifxetex 1\fi\ifluatex 1\fi=0 
  \usepackage[utf8]{inputenc}
\else 
  \usepackage{fontspec}
  \ifxetex
    \usepackage{xltxtra,xunicode}
  \fi
  \defaultfontfeatures{Mapping=tex-text,Scale=MatchLowercase}
  
\fi
\IfFileExists{microtype.sty}{\usepackage{microtype}}{}
\usepackage{graphicx}
\ifxetex
  \usepackage[setpagesize=false, 
              unicode=false, 
              xetex]{hyperref}
\else
  \usepackage[unicode=true]{hyperref}
\fi
\hypersetup{breaklinks=true,
            bookmarks=true,
            pdfauthor={},
            pdftitle={The Forecast Trap},
            colorlinks=false,
            urlcolor=blue,
            linkcolor=magenta,
            pdfborder={0 0 0}}
\newlength{\cslhangindent}
\newlength{\csllabelwidth}
  {}%
  {\par}
\newenvironment{CSLReferences}[2] 
 {
  \setlength{\parindent}{0pt}
  \ifodd #1 \everypar{\setlength{\hangindent}{\cslhangindent}}\ignorespaces\fi
  \ifnum #2 > 0
  \setlength{\parskip}{#2\baselineskip}
  \fi
 }%
 {}
\usepackage{calc}

\begin{document}
\begin{frontmatter}

  \title{The Forecast Trap}
    \author[a]{Carl Boettiger}
   \ead{cboettig@berkeley.edu} 
      \address[a]{Department of Environmental Science, Policy, and
Management, University of California, 130 Mulford Hall Berkeley, CA
94720-3114, USA}
    
  \begin{abstract}
  Encouraged by decision makers' appetite for future information on
  topics ranging from elections to pandemics, and enabled by the
  explosion of data and computational methods, model based forecasts
  have garnered increasing influence on a breadth of decisions in modern
  society. Using several classic examples from fisheries management, I
  demonstrate that selecting the model or models that produce the most
  accurate and precise forecast (measured by statistical scores) can
  sometimes lead to worse outcomes (measured by real-world objectives).
  This can create a forecast trap, in which the outcomes such as fish
  biomass or economic yield decline while the manager becomes
  increasingly convinced that these actions are consistent with the best
  models and data available. The forecast trap is not unique to this
  example, but a fundamental consequence of non-uniqueness of models.
  Existing practices promoting a broader set of models are the best way
  to avoid the trap.
  \end{abstract}
 \end{frontmatter}

Global change issues are complex and outcomes are difficult to predict
(Clark \emph{et al.} 2001). To guide decisions in an uncertain world,
researchers and decision makers may consider a range of alternative
plausible models to better reflect what we do and do not know about the
processes involved (Polasky \emph{et al.} 2011). Forecasts or
predictions from possible models can indicate what outcomes are most
likely to result under what decisions or actions. This has made
model-based forecasts a cornerstone for scientifically based decision
making. By comparing outcomes predicted by a model to future
observations, a decision maker can not only \emph{plan} for the
uncertainty, but also \emph{learn} which models are most trustworthy.
The value of iterative learning has long been reflected in the theory of
adaptive management (Walters \& Hilborn 1978) as well as in actual
adaptive management practices such as Management Strategy Evaluation
(MSE) (Punt \emph{et al.} 2016) used in fisheries, and is a central
tenet of a rapidly growing interest in ecological forecasting (Dietze
\emph{et al.} 2018). But, do iterative learning approaches always lead
to better decisions?

In this paper, I demonstrate that the model that makes the better
prediction (defined as a strictly proper score, Gneiting \& Raftery
(2007)) is not necessarily the model that makes the better policy
(defined in terms of utility, e.g.~expected net present value, Clark
(1990)). I show that our best methods for learning about model structure
or parameters by repeatedly comparing forecasts to observations can be
counter-productive. Put another way, the value of information (VOI, as
measured by the expected utility given that information minus the
utility without it; see Howard (1966); Katz \emph{et al.} (1987)), can
actually be negative. When VOI is negative, the decision-maker may
become trapped into accepting mediocre outcomes derived from a model
that makes accurate forecasts, even when a less accurate model that
would generate better outcomes is available. This trap is invisible to
the manager unless sufficient alternative models outside the original
set are introduced. I will present two examples of this ``forecast
trap'' and examine how it arises as a result of non-uniqueness of models
(Oreskes \emph{et al.} 1994; Schindler \& Hilborn 2015) with respect to
either of these objectives.

The forecast trap is not the only mechanism by which some model-choice
methods lead to worse outcomes. Previous work has long acknowledged the
panoply of ways in which model-based decision making can go astray due
to conflicting incentives, implementation errors, or lack of resources
for monitoring and updating (e.g. Ludwig \emph{et al.} 1993). Another
widely recognized problem is that of over-fitting (Burnham \& Anderson
1998), in which the model that best fits historical data fails to best
predict future data (Ginzburg \& Jensen 2004). Under such circumstances,
it is easy to see how an over-fit model would also lead to bad outcomes.
However, over-fitting plays no role in the forecast trap, where model
predictions are assessed only using probabilistic forecasts, and not
observations which had previously been used to fit the models. Formally,
these scores satisfy the `proper scoring' rule of Gneiting \& Raftery
(2007), which proves no other probabilistic prediction \(Q(x)\) will
have a better expected score than that of the true model
(i.e.~generative process), \(P(x)\). Gneiting \& Raftery (2007)'s proof
of proper scoring has since become a critical tool to avoid over-fitting
when choosing models to make decisions, but as I illustrate, will not
prevent the forecast trap.

First, I will introduce a motivating example in which we will consider
two reasonable process-based models, A and B. Model A will produce very
accurate forecasts, but lead to much worse outcomes than Model B. Though
I will establish that these accurate forecasts in Model A are not the
result of chance or of over-fitting the data, this example may raise
more questions than answers. To get a better understanding of when the
forecast trap arises, I will turn to a simpler ecological model, to
which we may apply more sophisticated decision tools of iterative
forecasting and adaptive management. We will see that these approaches
do not avoid the forecast trap either. No collection of such examples
can establish precisely how common the forecast trap may be in
real-world applications. The examples do establish unequivocally that
achieving incrementally ever-more-accurate forecasts does \emph{not
guarantee} better decisions. I conclude by pointing to a range of
established and emerging approaches to quantitative decision-making
which are not based on forecasts. As sophisticated forecasting
techniques become more common-place in conservation and ecology, the
forecast trap is a reminder that we should not forget about these
alternatives.

\hypertarget{a-note-on-models-and-data}{%
\subsection{A note on models and data}\label{a-note-on-models-and-data}}

I will use the term ``model'' to refer to any set of equations or code
that can be used to produce a forecast. This term thus includes not only
process-based models, but could also statistical forecasting methods,
non-parametric approaches such as empirical dynamical modeling (Ye
\emph{et al.} 2015), or machine learning. Most such models must first be
calibrated to historical data before they can produce a forecast,
e.g.~by parameter fitting, expert knowledge, or some other means.
Different choices for those parameters create different forecasts, I
will refer to those different parameterizations as different models. It
is of course possible for a decision-maker to consider forecasts coming
from multiple structurally different models simultaneously, and
potentially assigning different weights to each model. As more data
becomes available, it is possible to update model parameters, or
equivalently, update the weights assigned across models. I will examine
such approaches for model ensembles and model updating further on.

I shall focus on examples involving fisheries management to illustrate
principles shared in many ecological systems. Fisheries are a
significant economic and conservation concern worldwide and their
management remains an important debate (e.g. Worm \emph{et al.} 2006,
2009; Costello \emph{et al.} 2016). Moreover, fisheries management has
been a proving grounds for theoretical and practical decision-making
issues (e.g. Clark 1973; Reed 1979; Walters 1981; Ludwig \& Walters
1982) arising in a wide range of other contexts, including invasive
species (Boettiger 2021), infectious diseases (Shea \emph{et al.} 2014;
Li \emph{et al.} 2017), fire management (Richards \emph{et al.} 1999)
conservation planning and prioritization (Wilson \emph{et al.} 2006;
Chadès \emph{et al.} 2008) climate policy (Nitzbon \emph{et al.} 2017)
and much else (Ludwig \emph{et al.} 1993; Lande \emph{et al.} 1994;
Polasky \emph{et al.} 2011).

In these examples, we will focus on situations in which our `data' comes
from a model simulation rather than empirical sources. Simulations are
simplifications of the real world -- just because a method works in a
simulation is no guarantee that it works in reality. Conversely, if
decision methods are not reliable even when applied to simulated cases,
we should be even more cautious in how we use them. Simulations also
allow us to consider many replicates and conduct experiments that would
be often impossible or unethical to perform in the real world: for
instance, does a given fishery experience better long-term outcomes when
managed according to forecasts derived from model 1 or from model 2?

\hypertarget{a-motivating-example}{%
\section{A motivating example}\label{a-motivating-example}}

To better understand how a model can produce a more accurate forecast
and yet still lead to a worse decision, it may be helpful to start with
a concrete example in which a manager faces a trade-off between
cormorant conservation and fish harvest. Fig 1 shows both the forecasts
and realized management outcomes of two alternative three-species
models, ``A'' and ``B'' (see Appendix A) in predicting the population
dynamics of striped bass (an economically important fishery) and
double-crested cormorants (a target species for conservation) which both
feed on a population of river herring (whose abundance we assume is not
measured). Model A accurately forecasts the abundance of both bass and
cormorants well into the future, but the optimal management strategy
derived from its forecasts leads to steadily declining species
abundances and overall disappointing net utility. Model B produces
substantially less accurate forecasts, but nevertheless achieves better
outcomes. The net utility under model B is in fact nearly identical to
the maximum utility attainable given the true model, while management
under model A achieves only 38\% of that utility.

This management problem is motivated by a real world example of a
herring fishery as described in Brias \& Munch (2021), in which our
manager seeks to balance multiple objectives of sustaining the cormorant
population while maximizing the economic value from harvesting both
herring and striped bass. In the scenarios depicted in Fig 1, I have
used a the richer five-species model introduced Brias \& Munch (2021) to
drive the underlying dynamics, which includes three competing species of
herring that are preyed upon by both the bass and the cormorants. Here,
the manager seeks to maximize utility given by the weighted sum of the
individual objectives (Brias \& Munch 2021), placing 50\% of the weight
on the conservation objective and splits the remaining weight evenly
over the harvests for predator (bass) and prey (herring) species. I have
assumed a partially observable system - in this case, the manager
measures only the abundance of bass and cormorant species, and not of
the three herring species. In this scenario, I have further assumed the
manager must choose a fixed fishing effort for herring and for the bass
harvest, I will consider more dynamic decision-processes later, but it
is worth noting that in many real world conservation settings policy
choices are highly constrained and frequent adjustment of those policies
may be costly or impossible. Likewise, the assumptions of partially
observable system and imperfect models are characteristic of ecological
decision-making. Equations and code for all models in this example are
presented in Appendix A.

Both models A \& B can be seen as alternative attempts to approximate
the ``true'' model (generative process), which in real systems is always
unknown and more complex than any model thereof. Model A assumes a
three-species Ricker model which closely matches the trophic structure
of the ``true'' five-species model, lumping the three competing herring
species into a single variable. Because herring abundance is not
observed directly, the model parameters related to herring growth are
less accurate than other parameters. Model B also lumps herring species
together into a single variable, but fails to reflect the trophic
relationship between bass and herring. Model B also oversimplifies the
relationship between cormorant and herring population. This does not
make Model B an unreasonable model out-of-hand -- all models contain
such simplifications (e.g.~our ``true'' model does not model the trophic
relationship between the herring and its food sources or environmental
conditions explicitly either). Both models are consistent with the
limited historical data available to them.

\begin{figure}
\centering
\includegraphics{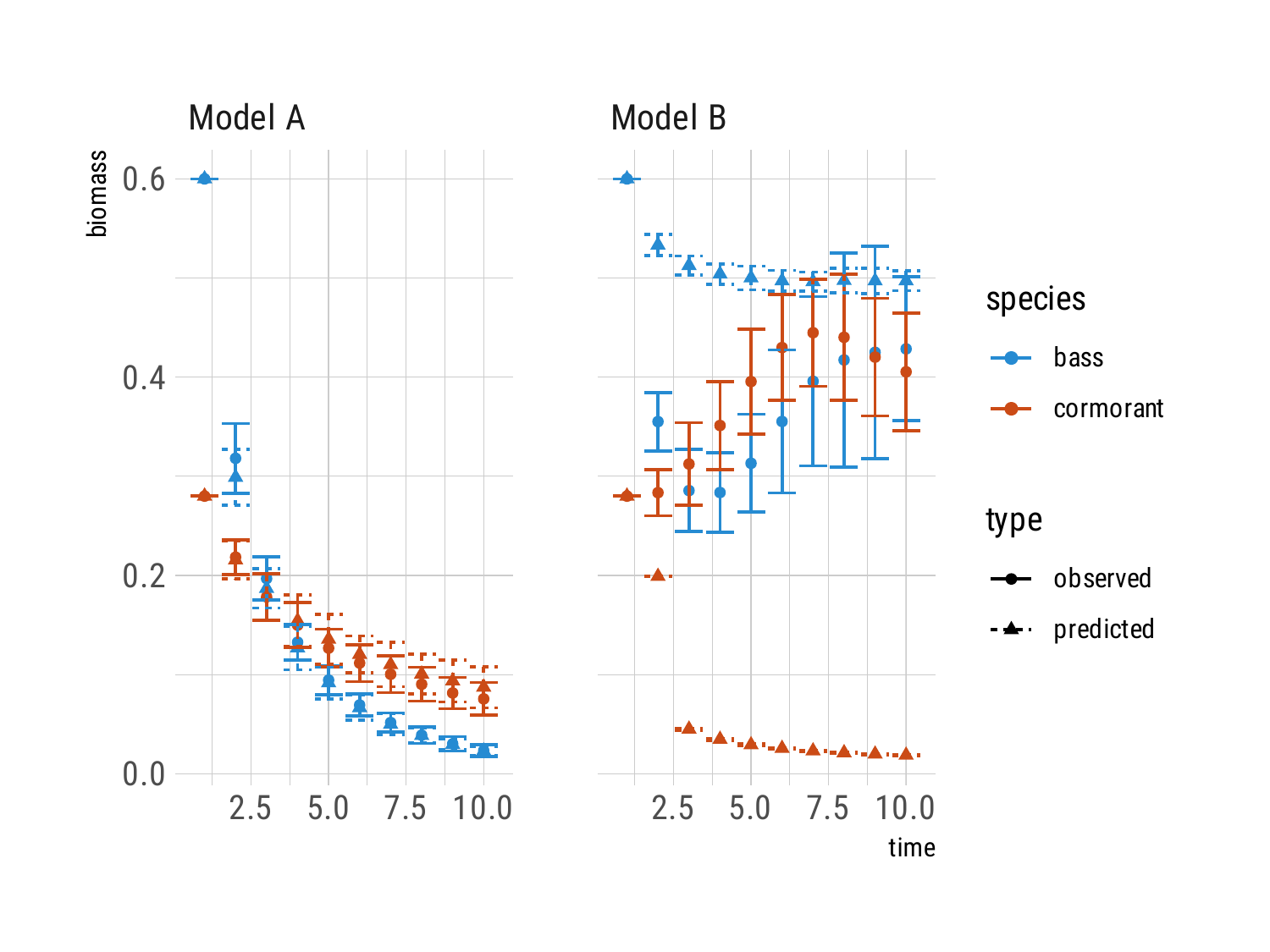}
\caption{Forecast performance and realized outcomes from management of a
5-species system under either model. Based on forecast performance
alone, Model A clearly performs better, accurately predicting steady
declines. Model B predicts overly optimistic outcomes for Bass
population levels, and overly pessimistic outcomes for cormorants, with
observed dynamics falling well outside the predicted range. Despite
this, net utility achieved under this management regime is virtually
optimal, while the declines under model A result in net utility that is
only 38\% of optimal.}
\end{figure}

\hypertarget{an-iterative-decision-example}{%
\section{An Iterative Decision
Example}\label{an-iterative-decision-example}}

While this example demonstrates that the model which provides the better
forecast does not necessarily lead to a better decision, it may raise
more questions than it answers. Why does this happen? Is this an
isolated example or not? Can this forecast trap be resolved by more
sophisticated approaches to model selection and decision-making? I now
consider scenarios involving iterative forecasts and adaptive
management: in which the manager monitors outcomes, compares forecasts
to observations and updates model estimates. Such \emph{sequential
decision processes} are not merely iterative versions of single-decision
problems, but are much more challenging. In the opening example, the
manager had to choose the harvest policy for each fish species at the
start of the scenario and stick with it. The ability to select new
actions in response to new observations turns that decision into a game
of chess: each turn, the manager must consider not only their next move
but all possible series of moves.

How do we translate a model-based forecast into a decision policy? It is
impossible to discuss outcomes associated with a forecast without first
agreeing on this process. In practice, decision-makers may use a
forecast in a wide variety of ways in selecting a course of action,
including ways which may run counter to the stated objectives of
management (Ludwig \emph{et al.} 1993). In principle at least, the field
of decision theory provides a formal mechanism for determining the
optimal strategy given a model forecast. For instance, a wide range of
ecological conservation and management problems can be expressed as a
Markov Decision Process (MDP) problems (Marescot \emph{et al.} 2013).
Existing computer algorithms such as stochastic dynamic programming
(SDP) take a probabilistic model \emph{forecast} (more precisely, the
probability \(P(x_{t+1} | x_t, a_t)\) of the system being in state
\(x_{t+1}\) in the next iteration given that it was previously in state
\(x_t\) and the manager selected action \(a_t\)) and the \emph{desired
management objective} (i.e.~the maximize the expected biomass of species
protected or the expected dollar profit of a fishery (see Clark 1990;
Halpern \emph{et al.} 2013)) as input, and return the \emph{decision
policy} which maximizes that objective (Marescot \emph{et al.} 2013).
This provides a principled way to associate a decision policy with any
given forecast model.

Two features of this approach are worth emphasizing. As before, the
resulting decision is derived directly from the forecast model and the
desired objective. The SDP algorithm is a reasonable description of the
approach any ideal manager would use -- considering all possible
outcomes from all possible sequences of actions and selecting the best
sequence. For complex models this process is too laborious even for a
computer, and is often simplified by considering only a selection of
predetermined policies (as in Management Strategy Evaluation, MSE, Punt
\emph{et al.} (2016)), or scenarios (as in scenario analysis, Polasky
\emph{et al.} (2011)). Such shortcuts are often necessary for complex
real-world models, but open additional room for error: the policy we
derive from a given forecast may perform poorly not because the model
forecast was at fault, but because of those simplifying assumptions
about possible policies. To ensure that the forecast trap is not a
result of such assumptions about possible policies, we will consider a
problem simple enough to solve directly with SDP. The resulting decision
policy is optimal, so long as the forecast model is correct. In this
way, the SDP merely stands in for a mathematically precise way in which
forecasts are turned into decisions. Recognizing the SDP-derived policy
(A) comes directly from the forecast model, and (B) gives the optimal
policy for said forecast, seems to suggest that whichever model makes
the better forecast will surely also lead to better outcomes (as
measured in terms of whatever utility we have chosen to maximize). While
this intuition is no doubt \emph{often} accurate, our purpose here is to
demonstrate that it is by no means \emph{guaranteed}: it is also
possible for the model which makes the better forecast to lead to worse
outcomes.

Let us consider the management of single species in which we seek to
maximize the long-term net harvest. In this scenario, the manager
estimates the population size each year and must set the total allowable
catch (TAC) for that season. The underlying dynamics are unknown, but
the manager is presented with any of a variety of forecast models which
can predict the future stock sizes given the current population size and
proposed TAC. Our manager does not know which of these models is the
most accurate a priori. Instead, the manager will be able to compare the
population size predicted by each forecast (under the chosen TAC) to the
measurement of the population size in the following year before coming
up with the next year's catch limit.

Faced with a collection of models, a manager can seek either to identify
the best model to use, or to consider an integrative assessment which
uses the whole ensemble of models to represent the manager's uncertainty
about the underlying process. I will consider both approaches in turn.
Figure 2 compares forecasts generated by two of the candidate models to
observations drawn from simulations of the underlying process. As
before, these are true forecasts: the model forecasts are generated
first, they have not been fit to these observations. In the un-fished
scenario (top panels), both models try to predicting the same un-fished
equilibrium dynamics. In the second scenario (lower panel), the manager
uses the optimal SDP policy derived from each forecast to determine the
TAC for the following year, and compares the observed stock size to that
which the model predicted given that fishing quota. In both cases, model
2 provides far more accurate forecasts, as seen in the error bars and
confirmed by the distribution of proper scores {[}Fig 2C-D; Gneiting \&
Raftery (2007){]}. Model and simulation details are provided in Appendix
B.

\begin{figure}
\centering
\includegraphics{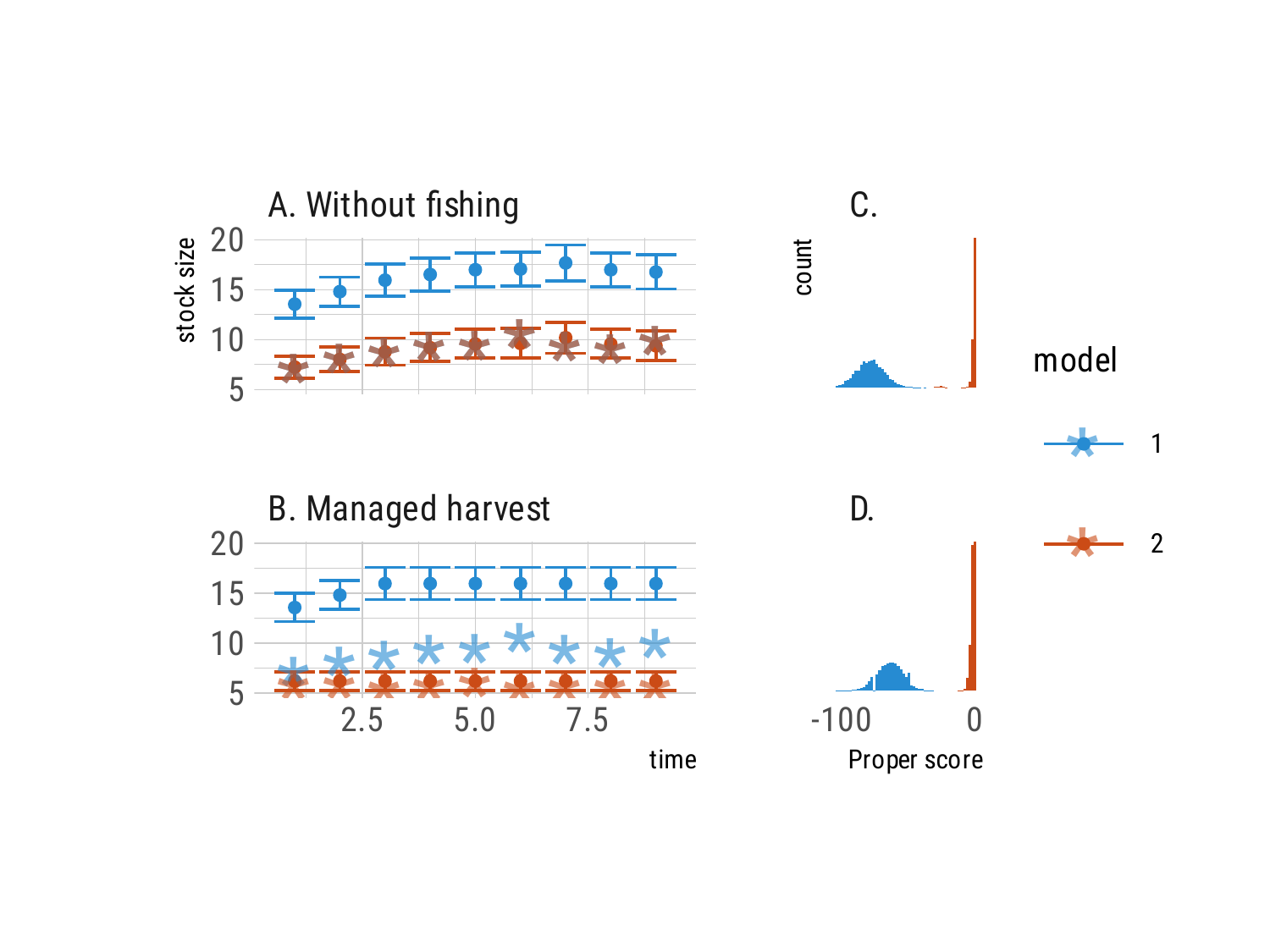}
\caption{Forecast performance of each model. Panels A, B: Step ahead
predictions of stock size under unfished (A) and fished (B) scenarios.
Error bars indicating the 95\% confidence intervals around each
prediction, while stars denote the observed value in that year. Because
the models make different decisions each year in the fished scenario,
the observed stock size in year 2, 3, etc under the management of model
1 (blue stars) is different from that under model 2 (red stars). Panels
C, D: corresponding distribution of proper scores across all predictions
(100 replicates of 100 timesteps). Higher scores are better, confirming
that model 2 makes the better forecasts.}
\end{figure}

Despite the clearly superior predictive accuracy of model 2 in both
scenarios, the outcomes from management under model 2 are substantially
worse. We can assess such outcomes in less abstract terms than
forecasting skill, such as economic value them manager sought to
optimize (in dollars) or the ecological value (unharvested biomass)
{[}Fig 3{]}.

\begin{figure}
\centering
\includegraphics{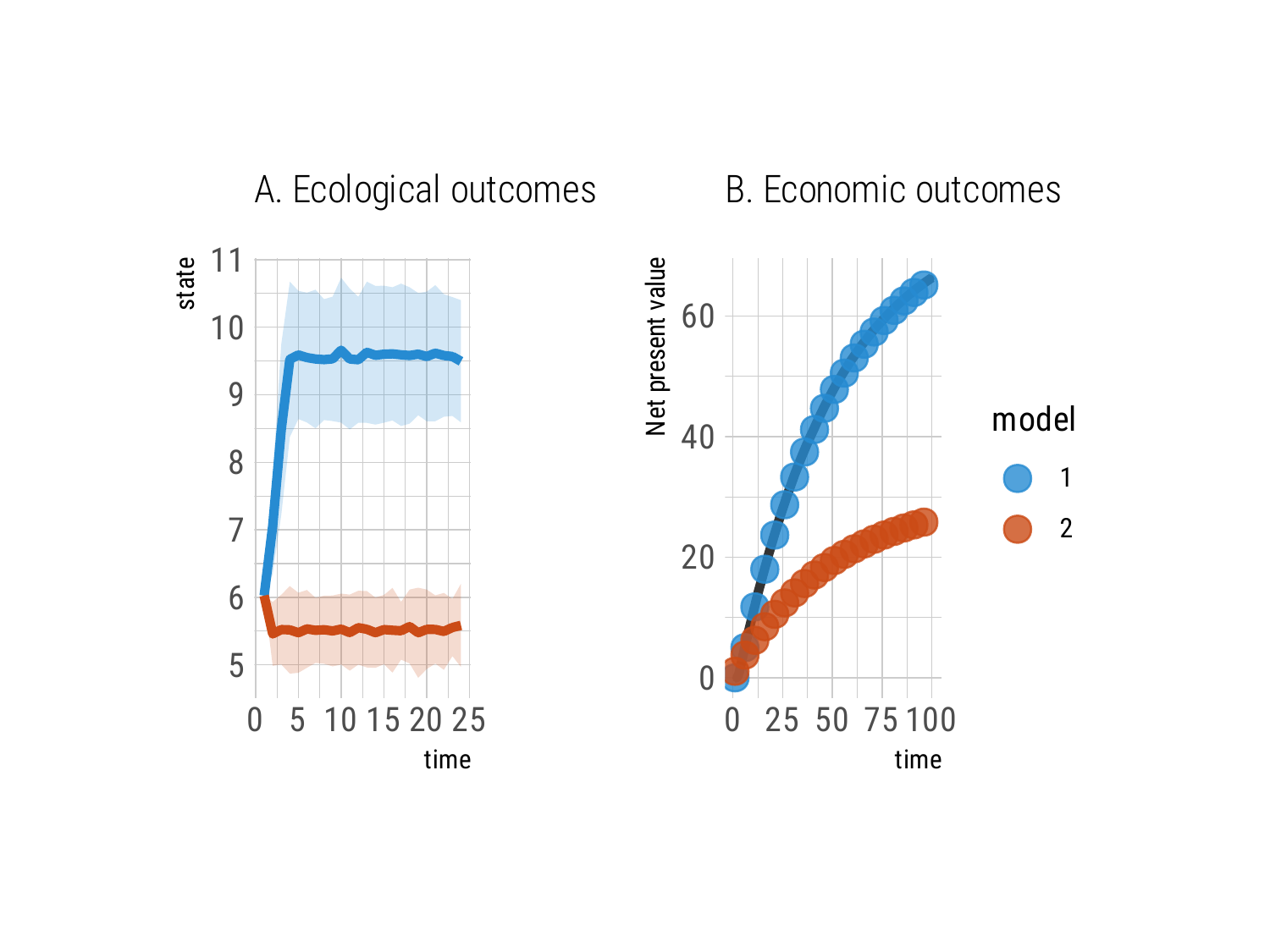}
\caption{Ecological and economic performance of each forecast. Harvest
quotas derived from model 1 result in a significantly higher fish stock
size than under Model 2 (panel A). Economic returns under model 1 are
also substantially higher (panel B)}
\end{figure}

A manager operating under model 2 would have little indication that the
model was flawed: both future stock sizes and expected harvest yields
consistently match model predictions. This manager would be stuck in the
forecast trap, incorrectly mistaking a degraded ecological state and
reduced economic outcomes as the best that can be achieved because
future observations continue to validate this model. Had we been able to
include Model 3 in our forecast comparisons, it would equal or
outperform the forecasting skill of both model 1 and model 2 (as
guaranteed by the theorem of Gneiting \& Raftery (2007)), while also
matching the economic utility of model 1 (as guaranteed by the theorem
of Reed (1979)). In practice, we never have access to the generating
model, so it is reasonable to expect model selection to determine the
better approximation. As we see here, the better approximation for
forecasting future states does not in fact lead to better outcomes.

\hypertarget{adaptive-management}{%
\subsection{Adaptive Management}\label{adaptive-management}}

Rather than select a single best model (or best parameter value), a
manager could choose to integrate over possible outcomes generated from
all candidate models. Updating posterior distributions over parameters
and/or weights assigned to different models are examples of this kind of
adaptive management (Ludwig \& Walters 1982; Punt \emph{et al.} 2016). I
illustrate the application of such an adaptive management strategy,
following classic examples for parameter (Ludwig \& Walters 1982) or
structural (Smith \& Walters 1981) model uncertainty. I first consider
only the same two models considered in the previous example. I later
consider a larger suite of 42 models, spanning the parameter space of
Gordon-Schaefer curves. To avoid failure to explore sufficiently, (see
exploration-exploitation trade-off, e.g. Walters 1981), I assign prior
belief of 99\% weight on the optimally performing model, model 1.\\
For comparison, I consider the baseline case in which the manager does
not update posterior distribution over which models/parameter values are
correct (the manager still chooses a new TAC after each observation, but
does not update their belief in the model, i.e.~does not \emph{learn}
over time). The difference between the performance with and without
learning is known as the ``Value of Information'' (Howard 1966). In both
2-model and 42-model scenarios, the value of information is strongly
negative. The 2-model case achieves a net present value to -58\% of the
value of having used model 1 alone {[}Fig 4{]}. Including all 42 models
reduces this to a value of -32\%. Both harvests and fish biomass remain
significantly lower under adaptive learning scenarios.

\begin{figure}
\centering
\includegraphics{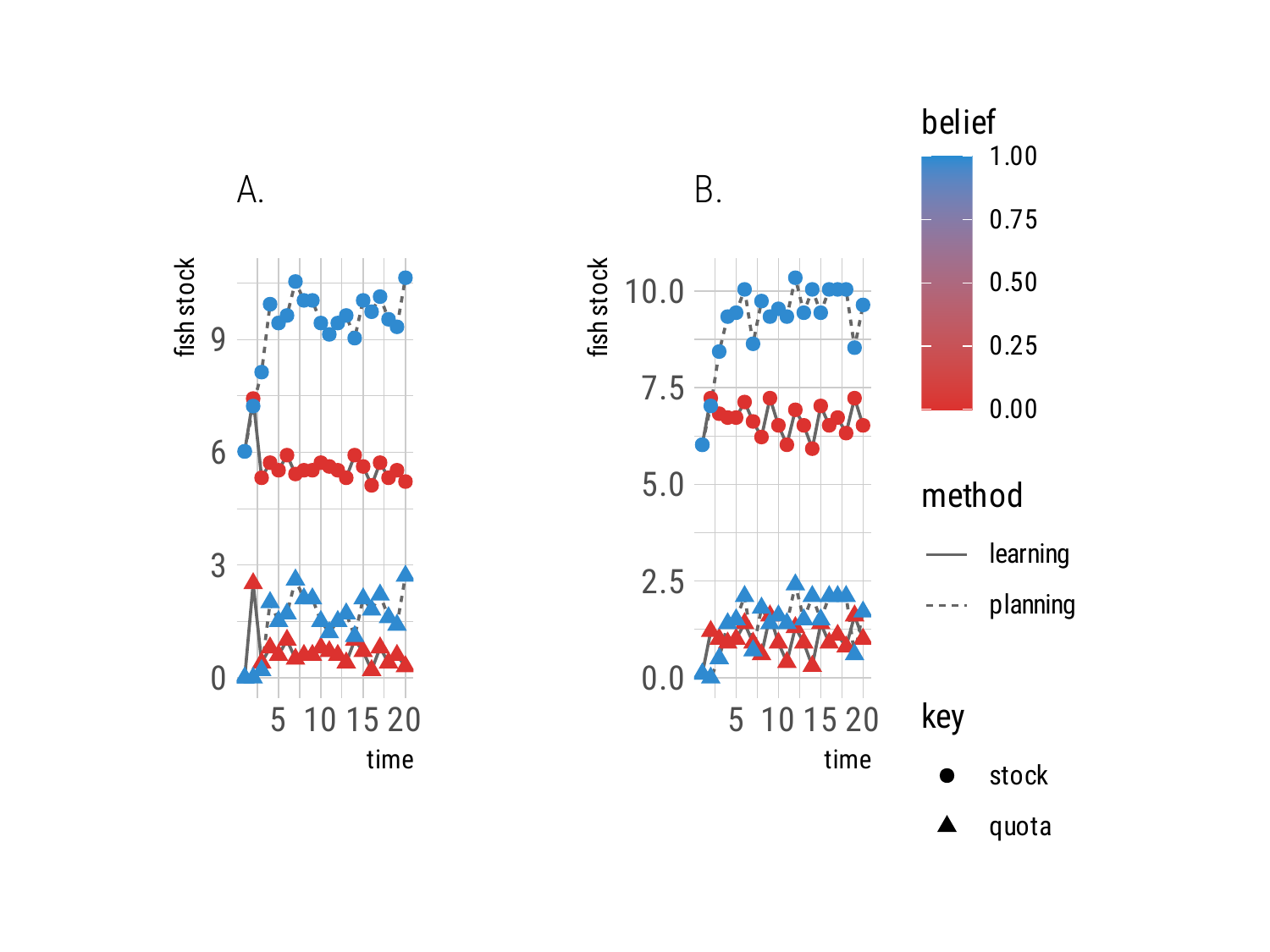}
\caption{Adaptive management under model uncertainty. Solid lines trace
the trajectories of the state (fish stock, circles) and action (harvest
quota, triangles), under adaptive management (learning). Dotted lines
trace the corresponding trajectories if iterative learning is omitted,
leaving the prior belief fixed throughout the simulation (planning).
Color indicates the belief that model 1 is correct (blue), with an
initial prior belief of 99\%. Panel A: Management over the two candidate
models, Model 1 and Model 2. Within a single iteration of adaptive
management, the belief over models switches from a prior belief that
heavily favored model 1 to a posterior that favors model 2 with near
certainty. Future iterations reinforce the belief in model 2, resulting
in both depressed harvests and low stock sizes (solid lines). If no
iterative learning updates are performed, stock sizes and realized
harvests (and thus economic profit) are both higher. Panel B: given 42
candidate models over a broad range of parameter values, adaptive
management quickly reduces the probability of model 1, and substantially
underperforms management without learning (dotted lines). While outcomes
improve marginally relative to the two-model case (panel A) they remain
significantly worse than had no iterative learning been included.}
\end{figure}

The reason for model 1's seemingly contradictory ability to make good
decisions but bad forecasts becomes obvious once we compare both curves
to that of the underlying model, model 3. Looking at plots of the growth
rate curves for each model {[}Fig 5A{]}, it is hardly surprising that
all model selection approaches prefer the closely overlapping curve of
model 2 to the no-where-close curve of model 1 as the better
approximation of model 3. Nevertheless, the decision policy derived from
model 1 forecasts is indistinguishable from that based on the true model
{[}Fig 5B{]}, while the policy derived from model 2 forecasts lead to
over-harvesting. Being closest to the true model's forecast skill never
guarantees that we are closest to the true model's optimal policy.

\begin{figure}
\centering
\includegraphics{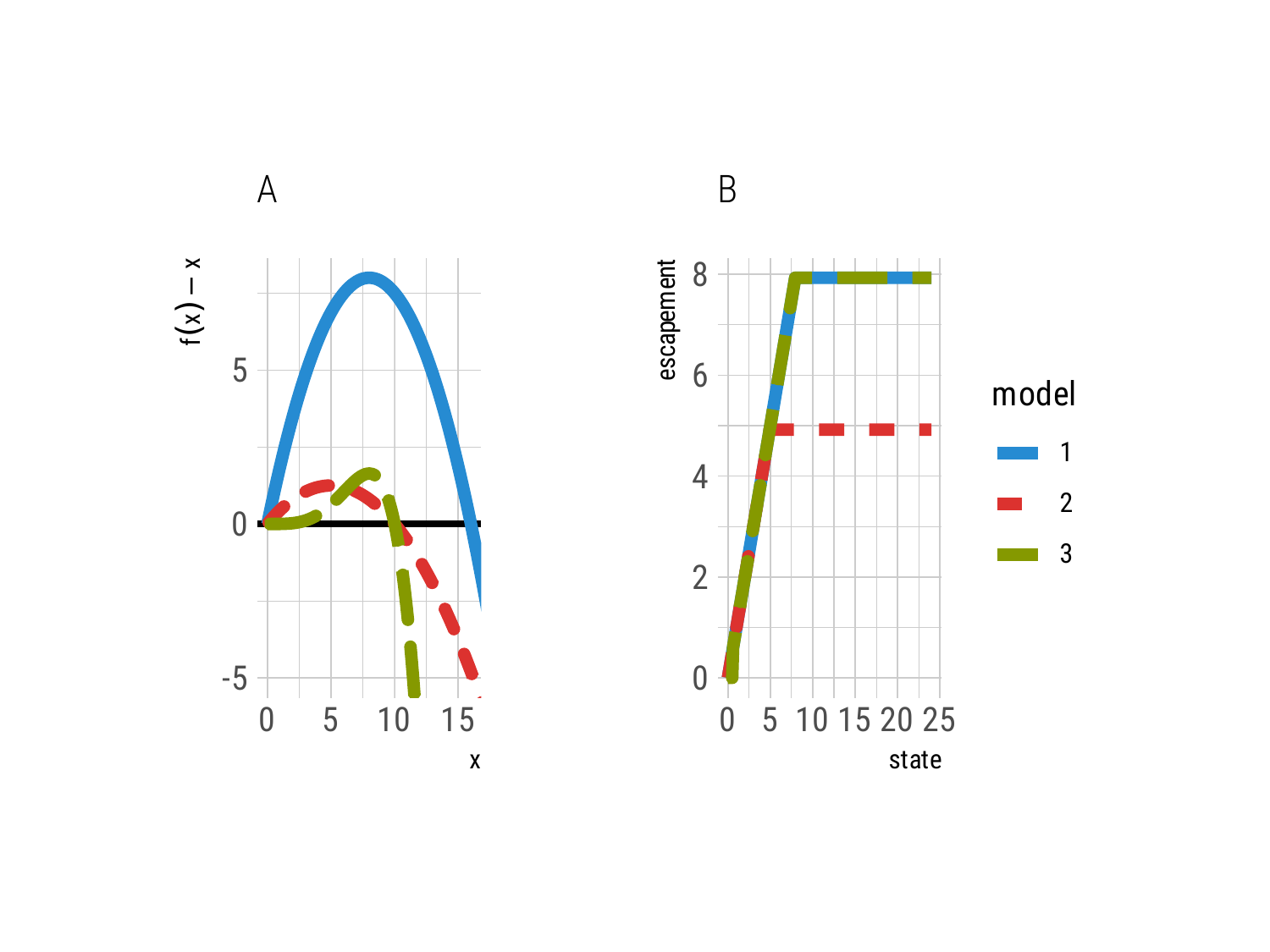}
\caption{Panel A: Population growth curves of each model. The positive
equilibrium of each model occurs where the curve crosses the horizontal
axis. Note that while Model 2 is a better approximation to the truth
(Model 3), Model 1 better approximates the stock size which leads to
maximum growth. Panel B: The optimal control policy under Model 1 is
nearly identical to that under the true Model 3, while the optimal
policy under Model 2 suppresses stock to a much lower escapement level.}
\end{figure}

How can the very different forecasts from model 1 and model 3 could
produce exactly the same optimal management policy (Fig 5B) under the
SDP algorithm? Analytic solutions offer more insight as to when and why
very different forecasts can generate the identical policy. Such a
solution was first provided by Reed (1979), who demonstrated the optimal
policy in the case considered here would be a so-called ``bang-bang''
policy. Intuitively one can think of this as maintaining the biomass at
the most productive size: the maximum population growth rate (position
of the peak of the growth curves in Fig 5A), though this is only
precisely true without discounting (\(\delta = 1\)): the optimal stock
size \(\hat x\) is the solution to \(f(\hat x) = \hat x/\delta\) when
stochasticity is sufficiently small (Reed 1979). Thus, all models in
which the peak growth rate occurs at the same stock size will have the
same optimal policy. These are not merely bad models getting lucky --
all such models correctly capture the crucial feature relevant to the
decision. In more complex models, such features are more difficult or
impossible to identify analytically; but this does not mean they do not
exist. For instance, Recent mathematical breakthroughs such as Holden \&
Conrad (2015), Hening \emph{et al.} (2019), and Hening (2021) have
proven that the optimal harvest control rule in age-structured and
predator-prey systems maintain similar bang-bang dynamics. This means
that the optimal policy of such very complex models will once again be
shared by infinite number of simpler models.

\hypertarget{discussion}{%
\section{Discussion}\label{discussion}}

The forecast trap illustrated in both examples can best be understood as
a problem of \emph{non-uniqueness} (Oreskes \emph{et al.} 1994). Even
modestly complex models can successfully predict the observed dynamics,
but for wrong mechanistic reasons (Schnute \& Richards 2001; Schindler
\& Hilborn 2015). In both examples, a decision-maker who accepts the
model which leads to very accurate predictions as the basis for their
decision-making winds up in the forecast trap: accepting poor ecological
and economic outcomes as the best possible option. The space of possible
models is infinitely large when measured against any a scalar metric
such as mean forecast skill or expected net utility. Perhaps it should
be no surprise then that many models will achieve the same policy
outcomes or achieve comparable predictive accuracy. Just as the forecast
skill is not unique, both examples also demonstrated that the optimal
policy is not unique to the ``true model'' -- many models will result in
the same policy and achieve the same outcomes; despite making very
different forecasts.

The forecast trap is likely to be more common in contexts in which
systems are more complex, partially observed, and available actions are
constrained -- all features which are particularly common to ecological
management and conservation. Because simplicity of the second example
allows analytic theory to reveal a precise explanation, it is tempting
to assume the trap is only a consequence of examining overly simple
models. In fact, the opposite is true. In the second example, the
``true'' model is simple enough to be covered by a suitable candidate
set of models (e.g.~a Gaussian Process, Boettiger \emph{et al.} (2015)),
which would resolve the trap given sufficient data. In reality, our
models never span the true process. Partially-observed systems increase
the space of possible models that achieve comparable predictive
accuracy. Constraints on action space such as adjustment costs
(Boettiger \emph{et al.} 2016) or piecewise-linear control rules (Punt
2010) increase the space of models which will result in the same policy.
Both of these aspects make the forecast trap easier to encounter in our
opening example.

The forecast trap demonstrates that for certain ensembles of candidate
models, the value of information (VOI) (Howard 1966; Katz \emph{et al.}
1987) can in fact be negative. Consequently, methods to select models or
re-estimate parameters can lead to worse outcomes than had these new
observations simply been ignored. Crucially, a manager implementing the
optimal policy from the most predictive model sees no indication that
their models are wrong -- the declining ecosystem and economic returns
observed under model A in the first example or under iterative learning
in the second example are completely consistent with and predicted by
the models. Only by winding back the clock, making decisions based over
the original uncertainty without learning (Fig 4), can we see that
better outcomes could have been achieved. The policy derived without
learning reflects greater uncertainty: it is thus more \emph{robust}.

\hypertarget{a-way-forward}{%
\subsection{A way forward}\label{a-way-forward}}

In practice, managers rarely rely only on forecasting skill to assess
models, nor determine policies directly from forecasts alone. In both
examples presented above, the forecast trap is most likely in
circumstances where the collection of candidate models is insufficiently
broad. Management practice tends to emphasize approaches which
\emph{broaden} rather than \emph{narrow down} this candidate set. This
reflects the view that ``the primary values of ecosystem models are as
heuristic tools for communication and for developing scenarios to
express uncertainties and test policies'' rather than as a source of
reliable forecasts (Schindler \& Hilborn 2015). Such practices include:

\begin{enumerate}
\def\labelenumi{(\alph{enumi})}
\tightlist
\item
  emphasizing a better articulation of uncertainty \emph{a priori};
\item
  active exploration of alternative policies can reveal when the model
  set is inadequate
\item
  methods for generating strategies that are robust to the sort of
  uncertainty described here.
\end{enumerate}

A central role of models is to help articulate uncertainty around
possible outcomes rather than make precise predictions. As Schindler \&
Hilborn (2015) notes, ``current approaches to verification and
validation of ecosystem models likely produce overly optimistic
impressions of the reliability of forecasts underlying management and
conservation prescriptions.'' Forecasts based on non-mechanistic models
such as empirical dynamical modeling (EDM, Ye \emph{et al.} 2015; Brias
\& Munch 2021) may help in articulating a broader ensemble of scenarios
(Boettiger \emph{et al.} 2015). In contrast, if such approaches are
selected solely on forecasting skill, they may increase the probability
of the forecast trap. The danger of an insufficiently broad model
ensemble is well understood in many disciplines which use \emph{scenario
planning} to assist policy in accounting for irreducible uncertainties
(Peterson \emph{et al.} 2003).

Second, a manager may also escape the forecast trap by exploring actions
which are never recommended from any of the available forecasts. For
example, a manager looking at the low stocks and poor harvests achieved
under model 2, could decide experimentally to reduce fishing quota. This
will allow the population to enter a range of state-space where
discrepancies between model 2 and observations are more obvious. Runge
\emph{et al.} (2016) describes how such ``double-loop learning'' can be
used to identify when the entire model set is inaccurate, a problem
which is not solved by ``single-loop'' adaptive-management in our
examples above. Schindler \& Hilborn (2015) also underscores the value
of flexible policies; rather than ``managing solely within the range of
past variation; active probing is usually needed,'' and contrasts this
to a typical interpretation of the `precautionary principle' often cited
as a reason to avoid exploratory actions. However, just because active
exploration can escape the forecast trap does not mean it is always a
good idea.

Third, managers may emphasize policy robustness over forecast skill
(Schindler \& Hilborn 2015). In many formal treatments, this is not
qualitatively different to the analysis considered here: a manager
simply chooses a different utility function, such as minimizing `regret'
rather than maximizing expected value (Polasky \emph{et al.} 2011). Such
approaches are just as vulnerable to bad outcomes (as defined by their
own utility functions) whenever models are selected only on the basis of
forecast skill. Alternative approaches may not seek any such
optimization, emphasize the viability (Aubin 1991) of possible policy
under constraints. In practice, robust design may emphasize acceptable
performance across the widest possible array of scenarios
(e.g.~candidate models). This acts more like a sensitivity analysis of
utility with respect to underlying assumptions, rather than an
optimization routine (Fischer \emph{et al.} 2009; e.g. Punt \emph{et
al.} 2016). Computationally, the former is much simpler, allowing
researchers to evaluate the performance of a policy on more complex
simulations for which calculating the optimal policy would be
prohibitively difficult.

Finally, it is worth noting that decisions do not need to be premised on
a forecast at all, but can be premised entirely on the basis of past
experience: `If the fish stock size has gone up, increase harvest
slightly, otherwise, decrease slightly.' Such a policy is not optimal,
but it is robust across a wide range of unimodal stock-recruitment
curves without ever estimating a predictive model. This is the basis of
so-called `model-free' reinforcement learning algorithms such as DQN
(Mnih \emph{et al.} 2015) and SAC (Haarnoja \emph{et al.} 2018), which
train deep neural networks to learn a policy without ever attempting to
predict future states of the underlying process. Training such a
artificial intelligence agents across a wide suite of simulations, a
process known as curriculum learning (Graves \emph{et al.} 2017), mimics
the scenario analyses and search for robust policies. Such approaches
have been used to train agents to play 2600 Atari console games at
superhuman ability (Mnih \emph{et al.} 2015), outperform race-car
drivers (Wurman \emph{et al.} 2022) and control nuclear fusion reactions
(Degrave \emph{et al.} 2022). Such approaches are also not yet well
understood, introducing new risks as well as new
possibilities(Dulac-Arnold \emph{et al.} 2019; Henderson \emph{et al.}
2019).

\hypertarget{acknowledgements}{%
\subsection{Acknowledgements}\label{acknowledgements}}

The author acknowledges support from NSF CAREER Award \#1942280 and
helpful discussions with Melissa Chapman. The author is also deeply
grateful to the Michael Runge, Chih-hao Hsieh, Antoine Brias, and other
anonymous reviewers whose detailed feedback and insights have greatly
shaped the paper and my own understanding of these issues.

\pagebreak

\hypertarget{references}{%
\section*{References}\label{references}}
\addcontentsline{toc}{section}{References}

\hypertarget{refs}{}
\begin{CSLReferences}{1}{0}
\leavevmode\vadjust pre{\hypertarget{ref-Aubin1991}{}}%
Aubin, J.P. (1991). \emph{Viability theory}. Systems \& control.
Birkhauser, Boston.

\leavevmode\vadjust pre{\hypertarget{ref-Boettiger2021}{}}%
Boettiger, C. (2021).
\href{https://doi.org/10.1007/s12080-020-00477-4}{Ecological management
of stochastic systems with long transients}. \emph{Theoretical Ecology},
14, 663--671.

\leavevmode\vadjust pre{\hypertarget{ref-Boettiger2016}{}}%
Boettiger, C., Bode, M., Sanchirico, J.N., LaRiviere, J., Hastings, A.
\& Armsworth, P.R. (2016).
\href{https://doi.org/10.1890/15-0236}{Optimal management of a
stochastically varying population when policy adjustment is costly}.
\emph{Ecological Applications}, 26, 808--817.

\leavevmode\vadjust pre{\hypertarget{ref-Boettiger2015}{}}%
Boettiger, C., Mangel, M. \& Munch, S. (2015).
\href{https://doi.org/10.1098/rspb.2014.1631}{Avoiding tipping points in
fisheries management through {Gaussian} process dynamic programming}.
\emph{Proceedings of the Royal Society B: Biological Sciences}, 282,
20141631--20141631.

\leavevmode\vadjust pre{\hypertarget{ref-Brias2021}{}}%
Brias, A. \& Munch, S.B. (2021).
\href{https://doi.org/10.1016/j.ecolmodel.2020.109423}{Ecosystem based
multi-species management using {Empirical} {Dynamic} {Programming}}.
\emph{Ecological Modelling}, 441, 109423.

\leavevmode\vadjust pre{\hypertarget{ref-Burnham1998}{}}%
Burnham, K.P. \& Anderson, D.R. (1998).
\href{https://doi.org/10.1007/978-1-4757-2917-7_3}{Practical {Use} of
the {Information}-{Theoretic} {Approach}}. In: \emph{Model {Selection}
and {Inference}}. Springer New York, New York, NY, pp. 75--117.

\leavevmode\vadjust pre{\hypertarget{ref-Chades2008}{}}%
Chadès, I., McDonald-Madden, E., McCarthy, M.a., Wintle, B., Linkie, M.
\& Possingham, H.P. (2008).
\href{https://doi.org/10.1073/pnas.0805265105}{When to stop managing or
surveying cryptic threatened species.} \emph{Proceedings of the National
Academy of Sciences}, 105, 13936--40.

\leavevmode\vadjust pre{\hypertarget{ref-Clark1973}{}}%
Clark, C.W. (1973). \href{https://doi.org/10.1086/260090}{Profit
maximization and the extinction of animal species}. \emph{Journal of
Political Economy}, 81, 950--961.

\leavevmode\vadjust pre{\hypertarget{ref-Clark1990}{}}%
Clark, C.W. (1990). \emph{{Mathematical Bioeconomics: The Optimal
Management of Renewable Resources, 2nd Edition}}. Wiley-Interscience.

\leavevmode\vadjust pre{\hypertarget{ref-Clark2001}{}}%
Clark, J.S., Carpenter, S.R., Barber, M., Collins, S., Dobson, A.,
Foley, J.A., \emph{et al.} (2001).
\href{https://doi.org/10.1126/science.293.5530.657}{Ecological
{Forecasts}: {An} {Emerging} {Imperative}}. \emph{Science}, 293,
657--660.

\leavevmode\vadjust pre{\hypertarget{ref-Costello2016}{}}%
Costello, C., Ovando, D., Clavelle, T., Strauss, C.K., Hilborn, R.,
Melnychuk, M.C., \emph{et al.} (2016).
\href{https://doi.org/10.1073/pnas.1520420113}{{Global fishery prospects
under contrasting management regimes}}. \emph{Proceedings of the
National Academy of Sciences}, 113, 5125--5129.

\leavevmode\vadjust pre{\hypertarget{ref-deepmind_fusion}{}}%
Degrave, J., Felici, F., Buchli, J., Neunert, M., Tracey, B., Carpanese,
F., \emph{et al.} (2022).
\href{https://doi.org/10.1038/s41586-021-04301-9}{Magnetic control of
tokamak plasmas through deep reinforcement learning}. \emph{Nature},
602, 414--419.

\leavevmode\vadjust pre{\hypertarget{ref-Dietze2018}{}}%
Dietze, M.C., Fox, A., Beck-Johnson, L.M., Betancourt, J.L., Hooten,
M.B., Jarnevich, C.S., \emph{et al.} (2018).
\href{https://doi.org/10.1073/pnas.1710231115}{Iterative near-term
ecological forecasting: {Needs}, opportunities, and challenges}.
\emph{Proceedings of the National Academy of Sciences}, 115, 1424--1432.

\leavevmode\vadjust pre{\hypertarget{ref-Dulac-Arnold_2019}{}}%
Dulac-Arnold, G., Mankowitz, D. \& Hester, T. (2019).
\href{http://arxiv.org/abs/1904.12901}{Challenges of {Real}-{World}
{Reinforcement} {Learning}}. \emph{arXiv:1904.12901 {[}cs, stat{]}}.

\leavevmode\vadjust pre{\hypertarget{ref-Fischer2009}{}}%
Fischer, J., Peterson, G.D., Gardner, T.A., Gordon, L.J., Fazey, I.,
Elmqvist, T., \emph{et al.} (2009).
\href{https://doi.org/10.1016/j.tree.2009.03.020}{Integrating resilience
thinking and optimisation for conservation.} \emph{Trends in ecology \&
evolution}, 24, 549--54.

\leavevmode\vadjust pre{\hypertarget{ref-Ginzburg2004}{}}%
Ginzburg, L.R. \& Jensen, C.X.J. (2004).
\href{https://doi.org/10.1016/j.tree.2003.11.004}{Rules of thumb for
judging ecological theories}. \emph{Trends in Ecology \& Evolution}, 19,
121--126.

\leavevmode\vadjust pre{\hypertarget{ref-Gneiting2007}{}}%
Gneiting, T. \& Raftery, A.E. (2007).
\href{https://doi.org/10.1198/016214506000001437}{Strictly {Proper}
{Scoring} {Rules}, {Prediction}, and {Estimation}}. \emph{Journal of the
American Statistical Association}, 102, 359--378.

\leavevmode\vadjust pre{\hypertarget{ref-curriculum_learning}{}}%
Graves, A., Bellemare, M.G., Menick, J., Munos, R. \& Kavukcuoglu, K.
(2017). \href{http://arxiv.org/abs/1704.03003}{Automated {Curriculum}
{Learning} for {Neural} {Networks}}. \emph{arXiv:1704.03003 {[}cs{]}}.

\leavevmode\vadjust pre{\hypertarget{ref-SAC}{}}%
Haarnoja, T., Zhou, A., Abbeel, P. \& Levine, S. (2018).
\href{http://arxiv.org/abs/1801.01290}{Soft {Actor}-{Critic}:
{Off}-{Policy} {Maximum} {Entropy} {Deep} {Reinforcement} {Learning}
with a {Stochastic} {Actor}}. \emph{arXiv:1801.01290 {[}cs, stat{]}}.

\leavevmode\vadjust pre{\hypertarget{ref-Halpern2013}{}}%
Halpern, B.S., Klein, C.J., Brown, C.J., Beger, M., Grantham, H.S.,
Mangubhai, S., \emph{et al.} (2013).
\href{https://doi.org/10.1073/pnas.1217689110}{Achieving the triple
bottom line in the face of inherent trade-offs among social equity,
economic return, and conservation.} \emph{Proceedings of the National
Academy of Sciences}, 110, 6229--34.

\leavevmode\vadjust pre{\hypertarget{ref-Henderson2019}{}}%
Henderson, P., Islam, R., Bachman, P., Pineau, J., Precup, D. \& Meger,
D. (2019). \href{http://arxiv.org/abs/1709.06560}{Deep {Reinforcement}
{Learning} that {Matters}}. \emph{arXiv:1709.06560 {[}cs, stat{]}}.

\leavevmode\vadjust pre{\hypertarget{ref-Hening2021}{}}%
Hening, A. (2021).
\href{https://doi.org/10.1007/s00332-020-09667-0}{Coexistence,
{Extinction}, and {Optimal} {Harvesting} in {Discrete}-{Time}
{Stochastic} {Population} {Models}}. \emph{Journal of Nonlinear
Science}, 31, 1.

\leavevmode\vadjust pre{\hypertarget{ref-Hening2019}{}}%
Hening, A., Nguyen, D.H., Ungureanu, S.C. \& Wong, T.K. (2019).
\href{https://doi.org/10.1007/s00285-018-1275-1}{Asymptotic harvesting
of populations in random environments}. \emph{Journal of Mathematical
Biology}, 78, 293--329.

\leavevmode\vadjust pre{\hypertarget{ref-Holden2015}{}}%
Holden, M.H. \& Conrad, J.M. (2015).
\href{http://dx.doi.org/10.1016/j.mbs.2015.08.021}{Optimal escapement in
stage-structured fisheries with environmental stochasticity}.
\emph{Mathematical Biosciences}, 269, 76--85.

\leavevmode\vadjust pre{\hypertarget{ref-Howard1966}{}}%
Howard, R. (1966).
\href{https://doi.org/10.1109/TSSC.1966.300074}{Information {Value}
{Theory}}. \emph{IEEE Transactions on Systems Science and Cybernetics},
2, 22--26.

\leavevmode\vadjust pre{\hypertarget{ref-Katz1987}{}}%
Katz, R.W., Brown, B.G. \& Murphy, A.H. (1987).
\href{https://doi.org/10.1002/for.3980060202}{Decision-analytic
assessment of the economic value of weather forecasts: {The}
fallowing/planting problem}. \emph{Journal of Forecasting}, 6, 77--89.

\leavevmode\vadjust pre{\hypertarget{ref-Lande1994}{}}%
Lande, R., Engen, S. \& Saether, B.-E. (1994).
\href{https://doi.org/10.1038/372088a0}{Optimal harvesting, economic
discounting and extinction risk in fluctuating populations}.
\emph{Nature}, 372, 88--90.

\leavevmode\vadjust pre{\hypertarget{ref-Li2017}{}}%
Li, S.-L., Bjørnstad, O.N., Ferrari, M.J., Mummah, R., Runge, M.C.,
Fonnesbeck, C.J., \emph{et al.} (2017).
\href{https://doi.org/10.1073/pnas.1617482114}{Essential information:
{Uncertainty} and optimal control of {Ebola} outbreaks}.
\emph{Proceedings of the National Academy of Sciences}, 114, 5659--5664.

\leavevmode\vadjust pre{\hypertarget{ref-Ludwig1993}{}}%
Ludwig, D., Hilborn, R. \& Walters, C. (1993).
\href{https://doi.org/10.1126/science.260.5104.17}{Uncertainty,
{Resource} {Exploitation}, and {Conservation}: {Lessons} from
{History}}. \emph{Science}, 260, 17--36.

\leavevmode\vadjust pre{\hypertarget{ref-Ludwig1982}{}}%
Ludwig, D. \& Walters, C.J. (1982).
\href{https://doi.org/10.1016/0304-3800(82)90023-0}{Optimal harvesting
with imprecise parameter estimates}. \emph{Ecological Modelling}, 14,
273--292.

\leavevmode\vadjust pre{\hypertarget{ref-Marescot2013}{}}%
Marescot, L., Chapron, G., Chadès, I., Fackler, P.L., Duchamp, C.,
Marboutin, E., \emph{et al.} (2013).
\href{https://doi.org/10.1111/2041-210X.12082}{Complex decisions made
simple: A primer on stochastic dynamic programming}. \emph{Methods in
Ecology and Evolution}, 4, 872--884.

\leavevmode\vadjust pre{\hypertarget{ref-DQN}{}}%
Mnih, V., Kavukcuoglu, K., Silver, D., Rusu, A.A., Veness, J.,
Bellemare, M.G., \emph{et al.} (2015).
\href{https://doi.org/10.1038/nature14236}{Human-level control through
deep reinforcement learning}. \emph{Nature}, 518, 529--533.

\leavevmode\vadjust pre{\hypertarget{ref-Nitzbon2017}{}}%
Nitzbon, J., Heitzig, J. \& Parlitz, U. (2017).
\href{https://doi.org/10.1088/1748-9326/aa7581}{Sustainability, collapse
and oscillations in a simple {World}-{Earth} model}. \emph{Environmental
Research Letters}, 12, 074020.

\leavevmode\vadjust pre{\hypertarget{ref-Oreskes1994}{}}%
Oreskes, N., Shrader-Frechette, K. \& Belitz, K. (1994).
\href{https://doi.org/10.1126/science.263.5147.641}{Verification,
{Validation}, and {Confirmation} of {Numerical} {Models} in the {Earth}
{Sciences}}. \emph{Science}, 263, 641--646.

\leavevmode\vadjust pre{\hypertarget{ref-Peterson2003}{}}%
Peterson, G.D., Cumming, G.S. \& Carpenter, S.R. (2003).
\href{https://doi.org/10.1046/j.1523-1739.2003.01491.x}{Scenario
{Planning}: A {Tool} for {Conservation} in an {Uncertain} {World}}.
\emph{Conservation Biology}, 17, 358--366.

\leavevmode\vadjust pre{\hypertarget{ref-Polasky2011}{}}%
Polasky, S., Carpenter, S.R., Folke, C. \& Keeler, B. (2011).
\href{https://doi.org/10.1016/j.tree.2011.04.007}{{Decision-making under
great uncertainty: environmental management in an era of global
change}}. \emph{Trends in Ecology {\&} Evolution}, 26, 398--404.

\leavevmode\vadjust pre{\hypertarget{ref-Punt2010}{}}%
Punt, A.E. (2010). Harvest control rules and fisheries management. In:
\emph{Handbook of {Marine} {Fisheries} {Conservation} and {Management}
.} (eds. Grafton, RQ, Hilborn, R, Squires, D, Tait, M \& Williams, M).
Oxford University Press.

\leavevmode\vadjust pre{\hypertarget{ref-Punt2016}{}}%
Punt, A.E., Butterworth, D.S., Moor, C.L. de, De Oliveira, J.A.A. \&
Haddon, M. (2016). \href{https://doi.org/10.1111/faf.12104}{Management
strategy evaluation: Best practices}. \emph{Fish and Fisheries}, 17,
303--334.

\leavevmode\vadjust pre{\hypertarget{ref-Reed1979}{}}%
Reed, W.J. (1979).
\href{https://doi.org/10.1016/0095-0696(79)90014-7}{{Optimal escapement
levels in stochastic and deterministic harvesting models}}.
\emph{Journal of Environmental Economics and Management}, 6, 350--363.

\leavevmode\vadjust pre{\hypertarget{ref-Richards1999}{}}%
Richards, S.A., Possingham, H.P. \& Tizard, J. (1999).
\href{https://doi.org/10.1890/1051-0761(1999)009\%5B0880:OFMFMC\%5D2.0.CO;2}{Optimal
{Fire} {Management} for {Maintaining} {Community} {Diversity}}.
\emph{Ecological Applications}, 9, 880--892.

\leavevmode\vadjust pre{\hypertarget{ref-Runge2016}{}}%
Runge, M.C., Stroeve, J.C., Barrett, A.P. \& McDonald-Madden, E. (2016).
\href{https://doi.org/10.1038/nclimate3041}{Detecting failure of climate
predictions}. \emph{Nature Climate Change}, 6, 861--864.

\leavevmode\vadjust pre{\hypertarget{ref-Schindler2015}{}}%
Schindler, D.E. \& Hilborn, R. (2015).
\href{https://doi.org/10.1126/science.1261824}{Prediction, precaution,
and policy under global change}. \emph{Science}, 347, 953--954.

\leavevmode\vadjust pre{\hypertarget{ref-Schnute2001}{}}%
Schnute, J.T. \& Richards, L.J. (2001).
\href{https://doi.org/10.1139/f00-150}{Use and abuse of fishery models}.
\emph{Canadian Journal of Fisheries and Aquatic Sciences}, 58, 10--17.

\leavevmode\vadjust pre{\hypertarget{ref-Shea2014}{}}%
Shea, K., Tildesley, M.J., Runge, M.C., Fonnesbeck, C.J. \& Ferrari,
M.J. (2014).
\href{https://doi.org/10.1371/journal.pbio.1001970}{Adaptive
{Management} and the {Value} of {Information}: {Learning} {Via}
{Intervention} in {Epidemiology}}. \emph{PLoS Biology}, 12, 9--12.

\leavevmode\vadjust pre{\hypertarget{ref-Smith1981}{}}%
Smith, A.D.M. \& Walters, C.J. (1981).
\href{https://doi.org/10.1139/f81-092}{Adaptive {Management} of
{Stock}--{Recruitment} {Systems}}. \emph{Canadian Journal of Fisheries
and Aquatic Sciences}, 38, 690--703.

\leavevmode\vadjust pre{\hypertarget{ref-Walters1981}{}}%
Walters, C.J. (1981). \href{https://doi.org/10.1139/f81-091}{Optimum
{Escapements} in the {Face} of {Alternative} {Recruitment}
{Hypotheses}}. \emph{Canadian Journal of Fisheries and Aquatic
Sciences}, 38, 678--689.

\leavevmode\vadjust pre{\hypertarget{ref-Walters1978}{}}%
Walters, C.J. \& Hilborn, R. (1978).
\href{https://doi.org/10.1146/annurev.es.09.110178.001105}{Ecological
{Optimization} and {Adaptive} {Management}}. \emph{Annual Review of
Ecology and Systematics}, 9, 157--188.

\leavevmode\vadjust pre{\hypertarget{ref-Wilson2006}{}}%
Wilson, K.A., McBride, M.F., Bode, M. \& Possingham, H.P. (2006).
\href{https://doi.org/10.1038/nature04366}{Prioritizing global
conservation efforts}. \emph{Nature}, 440, 337--340.

\leavevmode\vadjust pre{\hypertarget{ref-Worm2006}{}}%
Worm, B., Barbier, E.B., Beaumont, N., Duffy, J.E., Folke, C., Halpern,
B.S., \emph{et al.} (2006).
\href{https://doi.org/10.1126/science.1132294}{{Impacts of biodiversity
loss on ocean ecosystem services.}} \emph{Science (New York, N.Y.)},
314, 787--90.

\leavevmode\vadjust pre{\hypertarget{ref-Worm2009}{}}%
Worm, B., Hilborn, R., Baum, J.K., Branch, T.A., Collie, J.S., Costello,
C., \emph{et al.} (2009).
\href{https://doi.org/10.1126/science.1173146}{{Rebuilding global
fisheries.}} \emph{Science (New York, N.Y.)}, 325, 578--85.

\leavevmode\vadjust pre{\hypertarget{ref-sony}{}}%
Wurman, P.R., Barrett, S., Kawamoto, K., MacGlashan, J., Subramanian,
K., Walsh, T.J., \emph{et al.} (2022).
\href{https://doi.org/10.1038/s41586-021-04357-7}{Outracing champion
{Gran} {Turismo} drivers with deep reinforcement learning}.
\emph{Nature}, 602, 223--228.

\leavevmode\vadjust pre{\hypertarget{ref-Ye2015}{}}%
Ye, H., Beamish, R.J., Glaser, S.M., Grant, S.C.H., Hsieh, C., Richards,
L.J., \emph{et al.} (2015).
\href{https://doi.org/10.1073/pnas.1417063112}{Equation-free mechanistic
ecosystem forecasting using empirical dynamic modeling}.
\emph{Proceedings of the National Academy of Sciences}, 112,
E1569--E1576.

\end{CSLReferences}

\end{document}